\documentstyle[12pt]{article}
\begin{document}
\begin{flushright}
LAEFF-96/01\\
January 1996\\

\end{flushright}
\vspace{.5cm}
\begin{center}
{\LARGE Black Hole in Thermal Equilibrium with a Spin-2 Quantum Field}
\end{center}
\vspace{1cm}
\begin{center}
{\large David Hochberg\footnote{Electronic address:
hochberg@laeff.esa.es}}\\
{\large \sl Laboratorio de Astrof\'isica Espacial y F\'isica Fundamental\\
Apartado 50727, 28080  Madrid, Spain}
\end{center}
\begin{center}
{\large Sergey V. Sushkov\footnote{Electronic address:
sushkov@kspu.ksu.ras.ru}} \\
{\large \sl Department of Geometry, Kazan State Pedagogical
University\\ Mezhlauk str., 1, Kazan 420021, Russia}
\end{center}
\begin{abstract} 
An approximate form for the vacuum averaged stress-energy tensor of a
conformal spin-2 quantum field on a black hole background is employed
as a source term in the semiclassical Einstein equations. Analytic 
corrections to the Schwarzschild metric are obtained to first order in
$\epsilon = {\hbar}/M^2$, where $M$ denotes the mass of the black
hole. The approximate tensor possesses the exact trace anomaly and the 
proper asymptotic behavior at spatial infinity, is conserved with
respect to the background metric and is uniquely defined up to a free
parameter $\hat c_2$, which relates to the average quantum fluctuation
of the field at the horizon. We are able to determine and calculate
an explicit upper bound on $\hat c_2$ by requiring that the entropy due to the 
back-reaction be a positive increasing function in $r$. 
A lower bound for $\hat c_2$ can be established by requiring that the metric
perturbations be uniformly small throughout the region
$2M \leq r < r_o$, where $r_o$ is the radius of perturbative validity of the
modified metric.
Additional insight into the nature of the perturbed spacetime 
outside the black hole is provided by studying the effective 
potential for test particles in the vicinity of the horizon.  
\end{abstract}

\section{Introduction}

The physics of black holes provides a fertile ground in which the
confluence of gravitation, quantum mechanics and thermodynamics takes
place. Progress in our understanding of this confluence as well as of
the specific thermal and mechanical
aspects of black holes requires one to construct and study
model theories of semiclassical black holes which can provide insights
into the kinds of physical effects that may be present in a complete
and, as of yet, unrealized description of quantum gravity. 
A key to one such model
theory is the fact that a black hole can exist in (possibly unstable)
thermodynamic equilibrium provided it is coupled to thermal quantum
fields having a suitable distribution of stress-energy. In the
semiclassical approach, such fields are characterized by the vacuum
average of a stress-energy tensor obtained by the renormalization of a
quantum field on the classical background geometry of a black hole.
Using such a tensor as a source in the Einstein equation
\begin{equation}
G_{\mu \nu} = 8\pi <T_{\mu \nu}>_{ren}
\end{equation}
defines the associated semiclassical back-reaction problem. The solution of (1)
encodes the change induced by the stress-energy tensor on the black
hole's spacetime metric.

Before a solution can be obtained however, one needs to be able to
calculate the expectation value of stress-energy tensors for 
quantized fields in a suitable
vacuum state. This task has proven to be 
considerably difficult when the background spacetime is that
of a static black hole. Indeed, to date, the only exact numerical 
calculations of $<T_{\mu \nu}>_{ren}$ on this background 
have been carried out for the conformal
scalar \cite{Howard84} , the U(1) gauge boson \cite{Jensen89} , and most 
recently, for the non-conformal
scalar field \cite{Anders93}. In each of these cases, excellent analytic
approximations to the exact, numerically calculated tensors 
have been found, and these have been used, in turn, via
the solutions of (1), to
explore the thermodynamical and mechanical consequences of the
back-reaction of spin-0 and spin-1 quantum fields on a 
black hole \cite{York85}-\cite{Hoch95}. 
The case of a massless spin-$\frac{1}{2}$ fermion has also been investigated 
in Ref \cite{HKY93} based on an approximate stress-energy tensor.
In this way, one has been able to investigate the effects of quantized
matter on the geometry of black holes, in a case-by-case fashion, and
rather novel spin-dependent effects have been uncovered in the process \cite{HKY94}. While much has been learned from these studies,
it is also clear that any discussion of back-reaction in quantum field theory
in curved spacetime should include the effects of linearized
gravitons (whose spin $= 2$), which are expected to contribute to the
one-loop effective stress-energy tensor a term of the same order 
(or perhaps higher order) as
those coming from the lower spin fields. A knowledge of how gravitons
behave near the singularity at the center of a black hole is likely to
be crucial to our understanding of quantum gravity.
One would also like to obtain a self-consistent picture of the black hole
evaporation problem.
However, at the present time, the calculation of
such a tensor is confronted by complicated technical difficulties, the
solution of which shall require a reconciliation between gauge
invariance and renormalization \cite{Jensen95}. Namely, while a
complete set of solutions for the linear graviton field equations
exists only in the radiation gauge, explicit renormalization has been
implemented only in the deDonder gauge. 

Until such time as the technical difficulties associated with the
nonconformal, spin-2 linear graviton can be overcome, it is worthwhile to
obtain some idea of the magnitude of the back-reaction arising from
the spin-2 nature of the graviton. This, it turns out, can be achieved
provided we are willing to dispense altogether with the inherently nonconformal
graviton, and consider instead a conformal spin-2 quantum field. 
We hasten to point out that we are not suggesting to model the
graviton with a conformal field, rather, we seek a concrete means for
assessing the importance of spin-2 in the back-reaction on a black hole. 
Nevertheless, it might well turn out that some of the results 
of this calculation could serve as a guide as to what to 
expect in the technically complicated graviton case.
The back-reaction of the conformal spin-2 field can be calculated 
employing, for example, the
approximate stress-energy tensor ansatz constructed by 
Frolov and Zel'nikov \cite{Frolov87}  
valid for massless, conformal fields in any static spacetime. The
main idea of their approach is to approximate $<T_{\mu \nu}>_{ren}$ by
a tensor expanded in a basis containing the curvature tensor, the
Killing vector, and their covariant derivatives, up to some order. The
resulting tensor is covariantly conserved with respect to the
background metric, possesses the correct (and exact) trace anomaly, and
obeys certain important scaling relations and boundary conditions. 
For a static black hole in a vacuum, a unique 
(up to a parameter $\hat c_2$) tensor is singled out
which should provide a reasonable approximation to $<T_{\mu
\nu}>_{ren}$ in the Hartle-Hawking vacuum state. 

That there can arise important features depending solely on the spin
of the field coupling to the black hole
is supported by results of the back-reaction analyses 
presented in \cite{HK93}-\cite{HKY94} .
These do indeed indicate an important dependence on the 
spin of the quantum field. For
example, the energy density of the spin-1 vector boson near the black
hole horizon is roughly 120 times greater in magnitude than that 
of the conformal scalar \cite{HK93}. A calculation 
of the radial acceleration of a massive test
particle initially at rest just outside the horizon also manifests a
curious spin dependence. The acceleration is enhanced for the spin-0
scalar and for the spin-$\frac{1}{2}$ fermion, 
but can be reduced for the spin-1 boson \cite{HKY93}, for a
sufficiently 
large  number (or multiplicity) of U(1)fields. Spin dependence 
also shows up in the
effective potential for test particles in the vicinity of the black
hole leading to either an increase or decrease in the black hole's
capture cross-section \cite{HKY94}. 
  
In Section II we discuss the relevant features of the  
Frolov and Zel'nikov approximate
stress-energy tensor needed for the present calculation 
and calculate the spin dependent parameters
needed to apply it to the spin-2 field. The metric perturbations
resulting from using this tensor as a source in the semiclassical
Einstein equation are calculated in Section III. The way in which the
black hole mass is renormalized and how the remaining constant of
integration gets fixed by the thermodynamic boundary conditions is
reviewed briefly.
Both upper and lower bounds for $\hat c_2$ result
from requiring that the metric perturbations due to the back-reaction
be uniformly small over the entire range $2M \leq r < r_o$, where
$r_o$ is the radius of perturbative validity of the solutions of (1).
In Section IV, we compute the entropy $\Delta S$ by
which the back-reaction of the spin-2 field augments the
Bekenstein-Hawking entropy. By requiring that the field increases the
thermodynamic entropy of the system, we are able to put an upper bound on the
constant ${\hat c_2}$, which measures the magnitude of the quantum
fluctuations of the field at the horizon. This particular bound is much more
stringent than that coming from the condition of the ``smallness'' of the solutions. Combining the results from perturbative validity plus well-behaved
entropy, we obtain the double inequality
$-3080 < {\hat c_2} < -1366$.  
We also calculate $\Delta S$
for the lower spin conformal fields $(s=0, \frac{1}{2} ,1)$ and compare these
results to previous entropy calculations based on exact stress-energy tensors
in order to get some indication for the accuracy of the 
Frolov-Zel'nikov approximation. 
Further insight into the
nature of the modified spacetime geometry is obtained examining the
effective potential for test particle orbits in Section V. 
Our results are summarized briefly in the final Section. 
Units are chosen such that $G=c=k_B=1$ but ${\hbar} \neq 1$.

\section{Approximate Spin-2 Stress-Energy Tensor}

For the case of a static black hole background with metric $(w=2M/r)$
\begin{equation}
ds^2= -(1-w)dt^2 + (1-w)^{-1} dr^2 + r^2 (d\theta^2 + \sin^2 \theta \,
d\phi^2),
\end{equation}
the tensor ansatz constructed by Frolov and Zel'nikov \cite{Frolov87}  
takes on a relatively simple structure depending on 
just three spin-dependent constants  
\begin{equation}
T_{\mu \nu} = a_s\, T^{(tr)}_{\mu \nu} + b_s \tau'_{\mu \nu} + c_s\,
\tau''_{\mu \nu}.
\end{equation}
Defining the following constant tensors
\begin{eqnarray}
\Pi^{\nu}_{\mu} & = & \delta^{\nu}_{\mu} - 4\delta^0_{\mu}
\delta^{\nu}_0,\\
\Psi^{\nu}_{\mu} & = &  \delta^1_{\mu} \delta^{\nu}_1  -
\delta^0_{\mu}
\delta^{\nu}_0,
\end{eqnarray}
then the various terms in the expansion of $T^{\nu}_{\mu}$ are given by
\begin{eqnarray}
T^{(tr) \nu}_{\mu} & = & 48\kappa^4 \, w^6 (\delta^{\nu}_{\mu} + 
 3 \Pi^{\nu}_{\mu} - 6\Psi^{\nu}_{\mu})\\ 
\tau'^{\nu}_{\mu} & = &
\kappa^4[(1+2w+3w^2+4w^3+5w^4+6w^5-105w^6)\Pi^{\nu}_{\mu}\\ \nonumber
& + & 168w^6 \Psi^{\nu}_{\mu}], \\
\tau''^{\nu}_{\mu} &=& \kappa^4\,w^3[(4+5w+6w^2+15w^3)\Pi^{\nu}_{\mu}
\\ \nonumber
&-& 12(1+w+w^2+2w^3)\Psi^{\nu}_{\mu}],
\end{eqnarray}
where $\kappa = (4M)^{-1}$ is the surface gravity of the black hole. 
The constants in (3) are fixed from knowledge of the exact trace
anomaly and by boundary conditions to be satisfied 
at the black hole horizon and at
spatial infinity. All three tensors are finite at the horizon ($w=1)$.
The first tensor, $T^{(tr) \nu}_{\mu}$, is the only one with nonzero
trace:
\begin{equation}
T^{\mu}_{\mu} = a_s \,T^{(tr) \mu}_{\mu} = a_s\, 
\left( \frac{48M^2}{r^6} \right).
\end{equation}
On the other hand, the exact trace anomalies for conformal quantum
fields of arbitrary spin on a curved background 
have been calculated previously \cite{Capper74,Christens78}.
The general result can be expressed in terms of a certain linear
combination of curvature invariants \cite{Birrell82}.  
For the case of Ricci flat ($R_{\mu \nu} = 0$) 
backgrounds, the trace anomaly simplifies to
\begin{equation}
<T^{\mu}_{\mu}>_{ren} = \frac{\hat a_s}{(2880\pi^2)}\, R_{\alpha \beta
\gamma \delta}\, R^{\alpha \beta \gamma \delta} = \frac{\hat
 a_s}{(2880\pi^2)}\, \left( \frac{48M^2}{r^6} \right),
\end{equation}
where the final equality holds for the black hole background. In
particular, for spin-2, ${\hat a_2} = 212$, and matching coefficients
between (9) and (10) yields the identification $a_2 = \frac{212}{(2880\pi^2)}$.
The second tensor, $\tau'^{\nu}_{\mu}$, is the only one contributing
at far distances from the singularity $(r \rightarrow \infty)$, and takes on the asymptotic form,
\begin{equation}
T^{\nu}_{\mu} \rightarrow b_s\, \tau'^{\nu}_{\mu} = b_s\, \kappa^4 \,
{\rm diag}\left(-3,1,1,1 \right)^{\nu}_{\mu}.
\end{equation}
The coefficient $b_s$ is determined from the boundary condition that
{\it all} the quantum stress-energy tensors renormalized on a
black hole background 
approach the form of
a flat-spacetime radiation stress tensor at the uncorrected Hawking
temperature $(T_H = \frac{\kappa}{2\pi})$. 
The radiation stress tensor in flat space is simply proportional to a
constant tensor, the proportionality factor depending on the number of
independent helicity states of the field in question.
For the case of spin-2,
the flat-space limit is given by  
\begin{equation}
<T^{\nu}_{\mu}>_{ren} \rightarrow (\frac{\pi^2}{90}) h(2) T^4_H \,
{\rm diag}\left(-3,1,1,1 \right)^{\nu}_{\mu},
\end{equation}
where $h(2) = 2$ is the corresponding number of independent helicity
states (this is also the number of independent components of the
linear graviton in $3+1$ dimensions). 
Matching coefficients of (11) and (12) in this limit yields $b_2 =
\frac{4}{(2880\pi^2)}$. The third and final tensor is finite at the horizon and
vanishes asymptotically at infinity as $r^{-3}$. 
It is therefore presumably important
only for the intermediate zone near the black hole horizon.
Moreover, the coefficient $c_s$ multiplying it can
be fixed unambiguously by requiring that \cite{Frolov87}
\begin{equation}
T^{\nu}_{\mu}|_{w=1} = <T^{\nu}_{\mu}>^{ren}|_{w=1}.
\end{equation}
The tensor structure of both sides of this equation is identical at
the horizon, so only one constant is actually defined.
However, the implementation of this boundary condition
requires knowledge of the (exact) renormalized quantum stress
tensor at the black hole horizon. 
The exact value of $<T^{\nu}_{\mu}>^{ren}$ at the event horizon for
the Hartle-Hawking vacuum is known only for the conformal scalar
\cite{Howard84}, the non-conformal scalar \cite{Anders93} and the
electromagnetic field \cite{Jensen89}. 
As it is the spin-2 version of $<T^{\nu}_{\mu}>^{ren}$ 
we are interested in approximating, we shall need a different (but physically
equivalent) criterion for establishing the value (or bounds) of $c_2$. 
Loose bounds on $ c_2$ may be established by appealing to perturbation
theory, as will be made explicit in the next Section. However, an improved
upper bound results from exploiting the physical properties of the thermodynamic entropy.
We will indeed 
return to this important 
point later on when we come to discuss the amount by
which the spin-2 quantum field augments the thermodynamical entropy of
the black hole. In the meantime, we continue in what follows, keeping 
$ c_2 $ as a free parameter.  

\section{Metric Perturbations}

With the explicit components of the approximate spin-2 stress tensor in hand,
we may now proceed to solve the back-reaction equation (1) to first
order in $\epsilon = {\hbar}/M^2 < 1$. 
As $T^{\nu}_{\mu}$ is a function only of the
radial coordinate, the resulting metric perturbations will be static
and spherically symmetric. The most general metric satisfying these
conditions involves two independent radial functions, and may be 
written as \cite{York85,HKY93}
\begin{equation}
ds^2 = - \left( 1 - \frac{2m(r)}{r} \right) \,e^{2\psi(r)} \,dt^2 + 
 \left( 1 -
\frac{2m(r)}{r} \right)^{-1}\, dr^2 + r^2\, d\Omega^2.
\end{equation}
Then, the linear perturbations to the metric result from expanding 
the two metric functions in $\epsilon$ as 
\begin{eqnarray}
 e^{\psi (r)} & = & 1 + \epsilon {\bar \rho}(r),\\
 m(r) & = & M(1 + \epsilon {\bar \mu}(r) ),
\end{eqnarray}
and the functions $\bar \rho$ and $\bar \mu$ are solutions of the
(linearized) Einstein equations
\begin{eqnarray}
\frac{d{\bar \rho}}{dr} & = & -\frac{16 \pi M^2}{\epsilon w^3} \, 
 (1 - w)^{-1} \left[ T^r_r - T^t_t \right],\\
 \frac{d {\bar \mu}}{dr} & = & \frac{32 \pi M^2}{\epsilon w^4} \,
T^t_t.
\end{eqnarray}   
These follow directly from (1) after substituting (14) into the
Einstein tensor $G_{\mu \nu}$ and expanding both sides to
$O(\epsilon)$ (the stress tensor is itself $O(\hbar) = O(\epsilon)$). 
The solutions of these equations involve simple radial integrations
which for the present case, upon integrating up from the horizon $(1
\leq w)$, yield
\begin{eqnarray}
{\bar \mu}(w) = &-& \frac{1}{6K} \left( 32{\hat a_2}(w^3 - 1) +  {\hat
b_2}(-w^{-3} - 3w^{-2} -9w^{-1} +12\ln(w) \right.  \\ \nonumber 
 & + &  15w +9w^2 -49w^3 +38 )
+ \left. {\hat c_2}(3w + 3w^2 +7w^3 -13) \right) + C_0 K^{-1},
\end{eqnarray}
and
\begin{eqnarray}
{\bar \rho}(w) = & - & \frac{1}{3K} \left( {\hat b_2} (-\frac{1}{2} w^{-2} -
  3w^{-1} +6\ln(w) +10w +\frac{15}{2}w^2  \right. \\ \nonumber
 & + & 7w^3 -21 ) - 
 \left. {\hat c_2}(w^3 + \frac{3}{2}w^2 + 2w -\frac{9}{2}) \right) + k_0 K^{-1}.
\end{eqnarray}
The existence of the limit $\lim_{w \rightarrow 1^+} \frac{(T^r_r -
T^t_t)}{1 - w}$ has been used to render the integration of (17) trivial.
Here, $K = 3840\pi$, ${\hat c_2}/c_2 = {\hat b_2}/b_2 = {\hat a_2}/a_2 =
2880\pi^2$ and
$C_0$ and $k_0$ are constants of integration.
If we set $\mu = {\bar \mu} - C_0K^{-1}$, the mass function can be
re-written to $O(\epsilon)$ as
\begin{eqnarray}
m(w) & = & M[1 + \epsilon(\mu(w) + C_0K^{-1})]\\ \nonumber
     & = & M(1 + \epsilon C_0 K^{-1})[1 + \epsilon \mu(w)] \\
\nonumber
     & = & M_{ren}[1 + \epsilon \mu(w)],
\end{eqnarray}
so that the integration constant $C_0$ serves to renormalize 
the (bare) black hole mass, and we henceforth 
write $M \equiv M_{ren}$ in what follows, with
the tacit understanding that this stands for the 
physical black hole mass. 
The unknown quantities in the perturbed metric are reduced to a single
integration constant, $k_0$, which can be determined after suitable
boundary  conditions are imposed.

The necessity for imposing boundary conditions has been exhaustively
discussed in previous work \cite{York85,HKY93,HKY94}. 
We recapitulate briefly the main
points of that discussion here. In the first instance, asymptotic
flatness does not fix the value of $k_0$. To appreciate this point, 
it suffices to note that 
${\bar \rho}(r) \sim \frac{\hat b_2}{6K} (r/2M)^2$ for $r
\rightarrow \infty$.
Related to this limit is the fact
that the stress-energy tensors employed in (1) are asymptotically
constant (see (11) and (12)), thus the radiation in a 
sufficiently large spatial region
surrounding the black hole would collapse onto the hole and thereby
produce a larger one. This asymptotic constancy of the (renormalized)
stress tensors is, of course, not an artifact of any 
approximation or regularization scheme. 
It is simply the {\em physical} condition required in order that 
an observer at spatial
infinity measure the correct value of the Hawking temperature (of the
unperturbed black hole). As the perturbations grow (in $r$)
without bound,
it is therefore necessary to implant the system consisting of black hole
plus thermal quantum fields in a finite cavity with a wall radius $r_o
> 2M$. The allowed values for $r_o$ can be determined explicitly by requiring
that the metric perturbations remain uniformly small over a certain radial
domain. 
The boundary condition and the region outside the cavity wall
are really to be thought of as the ambient spacetime in which the
system (black hole $+$ radiation)
is embedded. We shall choose microcanonical boundary
conditions, specifying thus the total energy $E(r_o)$ at the cavity
wall, and match on an exterior metric of Schwarzschild form with an
effective mass $M^{*} = m(r_o)$. Defining $\rho$ by 
${\bar \rho} = \rho + k_o K^{-1}$, the
continuity of the metric across the
wall yields the relation $k_o = -K\rho(r_o)$. The spacetime geometry,
including the back reaction, is thus now completely specified for $r \leq
r_o$ by
\begin{equation}
ds^2 = -\left( 1 - \frac{2m(r)}{r} \right) [1 + 2\epsilon(\rho(r) -
\rho(r_o))]\,dt^2 + \left( 1 - \frac{2m(r)}{r} \right)^{-1}\, dr^2
+r^2 \, d\Omega^2,
\end{equation}
and for $r \geq r_o$ by
\begin{equation}
ds^2 = -\left( 1 - \frac{2m(r_o)}{r} \right)\, dt^2 + \left(1 -
\frac{2m(r_o)}{r} \right)^{-1} \,dr^2 + r^2 \, d\Omega^2.
\end{equation}

The parameter $r_o$ and $\epsilon$ must be chosen so that the corrections 
to the back-reaction remain suitably small. That is, we must ensure that the 
effect of $T^{\nu}_{\mu}$ is a perturbation of the Schwarzschild geometry.
This condition will be satisfied provided the proper (orthonormal frame)
perturbations, given by
\begin{equation}
h^r_r=w\frac{\mu(w)}{1-w},
\end{equation}
\begin{equation}
h^t_t=-2[\rho(w_o)-\rho(w)]-h^r_r
\end{equation}
with $w_o=2M/r_o$, obey 
\begin{equation}
\epsilon|h^{\alpha}_{\beta}|\equiv\delta<1.
\end{equation}
The cavity radius should be chosen so that
$r_o\le r_{asymp}$, where $r_{asymp}$ is the asymptotic radius
which is the maximum radius for which the metric perturbations remain
small.
Note that $h^t_t=-h^r_r$ if $r=r_o$.
Hereafter, we will take $\delta=\epsilon$ for illustrative purposes.
Hence $|h^{\alpha}_{\beta}|=1$. In addition, we shall use $r_o=r_{asymp}$ in
the following. Now, to obtain the asymptotic radius $r_{asymp}(=r_o)$
we go to the limit $r\to r_o$ with $r_o$ tending to a very large but
finite value. Taking the leading terms in $|h^r_r|$, we can write

\begin{equation}
\lim_{r\to r_o,r_o\to\mbox{\small large value}}|h^r_r|\sim\frac{2}{3K}
\left(\frac{r}{2M}\right)^2 +\frac{13}{6K}\hat c_2
\left(\frac{2M}{r}\right)=\left(\frac{\delta}{\epsilon}\right)=1.
\end{equation}
Solving this equation gives the asymptotic radius $r_{asymp}$.

In order to establish the bounds for $\hat c_2$ we must use the
requirement of ''smallness'' of the perturbations $|h^{\alpha}_{\beta}|$ in
the entire region of $r$, $2M\le r\le r_o$.
In particular, let us check
$|h^{\alpha}_{\beta}|$ at the horizon where $r=2M$ (or $w=1$). To obtain
the value of $h^r_r$ let us re-write the expression (19) for
$\mu(r)$ as follows
\begin{eqnarray}
\mu(w)&=&-\frac{1-w}{6K}\left(32\hat a_2(-w^2-w-1)+
\hat b_2(49w^2+40w+25 \right.  \\ \nonumber 
& - &  13w^{-1}- 4w^{-2}-w^{-3})  \\ \nonumber
&+& \left.\hat c_2(-7w^2-10w-13)\right)-\frac{1}{6K}12\hat b_2\ln(w).
\end{eqnarray}
Now using Eq.(24) we find that the value of $h^r_r$
at the horizon ($w=1$) is
\begin{equation}
\left. h^r_r\right|_{w=1} =   \frac{20016 +30\hat c_2}{72382}.
\end{equation}
Here, we have used the results that
\begin{equation}
\lim_{w \rightarrow 1}\frac{\mu(w)}{1-w} =
-\frac{1}{6K}(-96{\hat a_2}+96{\hat b_2}-30{\hat c_2})+\frac{12{\hat b_2}}{6K}.
\end{equation}
The ln-term gives a positive contribution.
The requirement that $\epsilon|h^r_r|=\delta<1$ (or $|h^r_r|=1$) yields the
following double-inequality
\begin{equation}
-3080 \leq  \hat c_2 \le 1746.
\end{equation}
Now we may substitute the value $\hat c_2= -3080$ into the equation
(27). It is easy to see that the second term in
Eq.(27), which contains $\hat c_2$, is much less than the first
one. So we may neglect the second term, and then the equation takes the form
\begin{equation}
\frac{2}{3K}\left(\frac{r}{2M}\right)^2=1.
\end{equation}
We have from (32) the following value of the asymptotic
radius: $r_o=260M$.

Now let us consider the value of $h^t_t$ at the horizon ($w=1$). We will
take $\hat c_2= -3080$
and $r_o=260M$. Leaving only the main terms in the
expression (25) for $h^t_t$ and using (20) and
(29) we obtain:
\begin{eqnarray}
h^t_t|_{w=1}&=&-2\rho(w_o)-h^r_r|_{w=1}\approx -2\rho(w_o)+1 \\ \nonumber
&\approx&-\frac{1}{3K}\hat b_2w_o^{-2} +\frac{9{\hat c_2}}{3K}   +1 
\\ \nonumber
&\approx& -1.64
\end{eqnarray}
Hence, in order that the quantity $\epsilon|h^t_t|$  be small at the
horizon, it is necessary that $\epsilon\le 0.6$.

\section{Thermodynamical Entropy}

The thermodynamical entropy $S$ of the black hole in thermal
equilibrium with the conformal spin-2 quantum field can be computed
following the method presented in Ref \cite{HKY93}. From the first law of
thermodynamics applied to slightly differing equilibrium systems 
\begin{equation}
dE = dQ  \,\,\, (dr=0, r\leq r_o)
\end{equation}
and so
\begin{equation}
dS = \frac{dQ}{T} = \frac{dE}{T} = \beta\, dE
\end{equation}
where $\beta$ is the inverse local temperature \cite{York85} ,
\begin{equation}
\beta(w) = \frac{8\pi M}{\hbar} \left[1 + \epsilon(\rho(w) - n_2
K^{-1} ) \right] \left(1 - \frac{2m(w)}{r} \right)^{1/2},
\end{equation}
and 
\begin{equation}
E(r) = r - r[g^{rr}(r)]^{1/2}
\end{equation}
is the quasilocal energy \cite{Brown93}, with $g^{rr}$ determined by (22).
Choosing $M$ and $r$ as independent variables and fixing $r$, we can
integrate (35) to obtain the total system entropy as a
distribution for $r < r_o$:
\begin{equation}
S = \frac{4\pi M^2}{\hbar} + \Delta S,
\end{equation}
where
\begin{equation}
\Delta S = 8\pi \int^w_1 \left[ {\tilde w}^{-1}(\rho - \mu) +
 \frac{\partial \mu}{\partial {\tilde w}} - n_2 K^{-1}{\tilde w}^{-1}
\right]\, d{\tilde w},
\end{equation} 
and 
\begin{equation}
n_2 = \left( \frac{\partial K\mu}{\partial w} \right)_{w = 1} =
 - 16{\hat a_2} + 14{\hat b_2} - 5{\hat c_2}.
\end{equation}
The quantity $\Delta S$ is therefore the amount by 
which the quantum field changes
the Bekenstein-Hawking entropy, $S_{BH} \equiv {4\pi M^2}/{\hbar}$,
through its back-reaction. $S$ is a function of $r$ and is the total system
 entropy (black hole plus radiation) contained within the region of radius $r$.
Working out the integral with the explicit forms for $\rho$ and $\mu$
calculated above, we find that 
\begin{eqnarray}
\Delta S/{8\pi} & = & (\frac{32{\hat a_2}}{6K})
[-\frac{2}{3}w^3 + 2\ln(w) + \frac{2}{3}] \\ \nonumber
& + & (\frac{\hat b_2}{6K})[\frac{4}{3}w^{-3} +4w^{-2} +12w^{-1}
 - 16\ln(w) -20w \\ \nonumber 
& - & 12w^2 +28w^3 -\frac{40}{3}] \\ \nonumber
& + & (\frac{\hat c_2}{6K}) [4w - 4w^3 + 8\ln(w) ].
\end{eqnarray}
As a general consequence of (39), the horizon is a local extremum with
respect to $r$ since
\begin{equation}
\left( \frac{\partial \Delta S}{\partial w} \right)_{M, w=1} =
8\pi \left[w^{-1}(\rho - \mu) + \frac{\partial \mu}{\partial w}
- n_2 w^{-1} K^{-1} \right]_{w = 1} = 0,
\end{equation}
as follows from the fact that $\rho(1) = \mu(1) = 0$, and the
definition
of $n_2$. On physical grounds, we demand that the horizon be a local
minimum to prevent the existence of a spherical shell of negative
entropy near $r = 2M$. The same physical criterion was employed
recently to establish limits in the range of the non-minimal
coupling constant between the
scalar field and the scalar curvature 
\cite{Anders94}.
Computing the second derivative of $\Delta S$
at the horizon yields
\begin{equation}
\frac{K}{8\pi} \left( \frac{\partial^2 \Delta S}{\partial w^2}
\right)_{w = 1} = -32{\hat a_2} + \frac{112}{3}{\hat b_2}
-\frac{16}{3}
{\hat c_2}.
\end{equation}
Substituting in the values of $\hat a_2 = 212$ and $\hat b_2 = 4$,
we find that the horizon will be a local minimum of the entropy   
if and only if
${\hat c_2} < -1241.5$.
Values of ${\hat c_2}$ satisfying
this inequality will only guarantee that $\Delta S \geq 0$ 
and increasing at the horizon. In fact, Eq.(41), because it 
contains a term in $\hat c_2$, cannot be non-negative 
and monotone increasing (as a function of $r$ for fixed $M$) for all 
values of $\hat c_2$. 
In order that $\Delta S$ be a positive and increasing
function everywhere (strictly speaking, for $2M \leq r < r_o$), 
we cannot allow $({\partial \Delta S}/{\partial
r})_M = 0$ for {\em any} value of $r > 2M$. That is, we do not allow
for a spherical layer of negative entropy at any value of 
$r$ beyond the horizon. 
Explicit calculation of $\Delta S$
indicates that 
the smallest value (in absolute magnitude) 
of $\hat c_2$ satifying this condition 
is ${\hat c_2}= -1366$, which occurs at $r \sim 5.7M$ 
see e.g., Figure 1. We plot the quantity $(3K/4\pi) \Delta S$ 
versus $w=2M/r$ in Fig. 1 for the values of ${\hat c_2}$ indicated there. Note that large $(r >> 2M)$ values of the radial 
coordinate correspond to small values of $w$
$(0 < w \leq 1)$. One sees in particular that $\Delta S$ is positive and increasing at the black hole horizon since ${\hat c_2} < -1241.5$ in all the cases displayed, but that there can (and do) arise local extrema away from the horizon depending on the precise value of ${\hat c_2}$. 
For example, as indicated in Fig. 1, the value ${\hat c_2} = -1336$ yields
a $\Delta S$ which increases from the horizon, only to vanish again at
$r \sim 7.3M$, after which it increases once more.
The entropy due to the spin-2 back-reaction will
therefore be a positive and monotonically increasing 
function of $r$, within the present approximation,
provided that ${\hat c_2} < -1366$. Putting this result together with the
perturbative bounds obtained earlier, we conclude that 
$-3080 < {\hat c_2} < -1366$, simultaneously guaranteeing well-behaved entropy as well as the perturbative validity of the solution.
Returning to the boundary condition in (13), we see that $\hat c_2$
gives a measure of the average quantum fluctuation of the spin-2 field
at the black hole horizon. We may thus employ the criterion
of physically well-behaved entropy and perturbative validity 
to establish bounds in the range of spin-2 field
fluctuations at the horizon. Working out (13) for $\mu =\nu = 0$ 
(the other components give no new information),
we obtain the following bounds for the energy density at the horizon:
\begin{equation}
25.1\, >\, \frac{\pi^2}{\kappa^4} <T^t_t>^{ren}_{w=1}\, >\, 7.3.
\end{equation}
This should be contrasted to the exact result 
\begin{equation}
\frac{\pi^2}{\kappa^4} <T^t_t>^{ren}_{w=1}\, =\, 0.63,
\end{equation}
for the spin-1 case and to the corresponding exact result
\begin{equation}
\frac{\pi^2}{\kappa^4} <T^t_t>^{ren}_{w=1}\, =\, 0.05
\end{equation}
for the spin-0 case. 
The averaged quantum fluctuations for the spin-2 field at the horizon are 
at least 100 times larger than for the scalar field case, and at least 10 times
greater than for the spin-1 case.  

Though in the present work we are primarily interested in the
physical properties of the spin-2 back reaction, it is of interest
to apply the Killing approximation (3) in order to evaluate $\Delta S$
arising from the back-reaction of some lower spin conformal fields. Indeed, as
the entropy corrections have been calculated independently for the
cases spin $s = 0, \frac{1}{2}, 1$ \cite{HKY93}, we have 
a means of checking the
accuracy of this approximation explicitly for the conformal scalar and
for the U(1) gauge boson, two cases where exact renormalized stress-energy
tensors are known. Stress tensors renormalized on a Schwarzschild
background have been obtained in exact form by Howard \cite{Howard84} for the
conformal scalar and by Jensen and Ottewill (JO) \cite{Jensen89}   
for the abelian vector
boson. In both cases, there also exist excellent analytic
approximations for the full numeric calculations. The analytic form
for the conformal scalar 
was first given by Page (P) \cite{Page82}. We use these analytic
results to calculate ${\hat c_s}$ by means of the horizon boundary
condition, Eq(13). For the massless spin-$\frac{1}{2}$ fermion, we 
have used the
tensor of Brown, Ottewill and Page (BOP) \cite{Brown86}, but the 
accuracy of their
approximation has not, to our knowledge, been checked against an exact
numerical calculation. We collect the various spin-dependent
coefficients needed to evaluate $\Delta S$ based on  the Frolov and
Zel'nikov (FZ)
approximate tensor, in Table 1.
\begin{table}[t]
\caption{Spin-dependent constants appearing in Eq. (3)}
\vspace{0.2cm}
\begin{center}
\begin{tabular}{|l|l|l|l|} \hline
${\rm spin}= s $ & $\hat a_s$ & ${\hat b_s}$ & ${\hat c_s}$ \\ \hline
0            &   1        &    2         &     0    \\
$\frac{1}{2}$ & $\frac{7}{4}$ & $\frac{7}{2}$ &  5    \\ 
 1           &  -13          &     4         & -8    \\
2            & 212   &   4   &  $-3080 < {\hat c_2} < -1366$ \\
\hline
\end{tabular}
\end{center}
\end{table}

The comparison of the (spin-dependent) entropies calculated from the
various stress-energy tensors  may be summarized compactly in the following
manner:
\begin{eqnarray}
\Delta S_{FZ}  & \equiv & \Delta S_P , \,\, ({\rm spin=0})\\
\Delta S_{FZ}  & \equiv & \Delta S_{BOP}, \, \, ({\rm spin=1/2}),\\
\Delta S_{FZ} - \Delta S_{JO} &=& 32(w^3 - w - 2\ln(w) ), ({\rm
spin=1}).
\end{eqnarray}
The explicit expressions for the corrections calculated 
from other stress-energy tensors are presented in Table
2; the results are taken from Ref \cite{HKY93}.
\begin{table}[t]
\caption{Spin-dependent entropy corrections $\Delta S$ based on other
 stress-energy tensors.}
\vspace{0.3cm}
\begin{center}
\begin{tabular}{|l|l|l|} \hline
${\rm spin}=s$ & Stress tensor & $\Delta S(w)$  \\ \hline
0 & $\rm P$ & $\frac{8\pi}{K}[\frac{52}{9}w^3 - 4w^2 - \frac{20}{3}w
+\frac{16}{3}\ln(w)$ \\
  &      &                    \\
  &      & $ + 4w^{-1} + \frac{4}{3}w^{-2} +\frac{4}{9}w^{-3} -
\frac{8}{9}]$ \\ 
    &     &                 \\
$\frac{1}{2}$ & {\rm BOP} &
$\frac{7}{8}\frac{8\pi}{K}[\frac{488}{63}w^3 - 8w^2 -\frac{200}{21}w +
\frac{128}{7}\ln(w)$ \\
  &    &               \\
  &   &  $ + 8w^{-1} +\frac{8}{3}w^{-2} +\frac{8}{9}w^{-3}
-\frac{16}{9}]$ \\
    &       &         \\
1 & ${\rm JO}$ & $\frac{8\pi}{K}[\frac{344}{9}w^3 -8w^2 +
\frac{40}{3}w - 96\ln(w)$ \\
   &   &                   \\ 
   &   & $+ 8w^{-1} +\frac{8}{3}w^{-2} +
\frac{8}{9}w^{-3} -\frac{496}{9}]$ \\
\hline
\end{tabular}
\end{center}
\end{table}

As discussed in Ref \cite{HKY93}, the corrections $\Delta S_P \geq 0$, $\Delta
S_{BOP} \geq 0$ and $\Delta S_{JO} \geq 0$ and are monotonically increasing 
for all $r \geq 2M$. Since
the difference $(\Delta S_{FZ} - \Delta S_{JO})$ vanishes at the
horizon and grows as $64 \ln(w)$ for large $r$, $\Delta S_{FZ}$ is
also a positive and increasing function of $r$ for spin-1. 
The condition that the horizon be a local minimum 
of the entropy, $\Delta S_{FZ}$ Eq.(41),
yields the inequalities 
$\hat c_0 < 8$, $\hat c_{\frac{1}{2}} < 14$ and $\hat c_1
< 106$, respectively, which are automatically satisfied by the values
displayed in Table 1., derived from the 
exact stress tensors for $s=0,1$ and by the value
derived from the approximate BOP tensor for the spinor case. 
In Fig. 2 we plot the various spin-dependent entropy corrections $(3K/4\pi)
\Delta S(w)$ whose functional forms are listed in Table 2.
We should point out that in the case of
the electromagnetic field, an exact value of $<T_{\mu}^{\nu}>_{ren}$ at the
black hole horizon was calculated some time 
ago in Refs \cite{Elster, Frolov85}. 
Those results lead, however, to a value for $\hat c_1 = 92h(2) = 184$, which
clearly violates the entropy positivity bound cited above. The reason for the 
large discrepancy between the two values of $\hat c_1$ (-8 versus 184) 
is due to an important linearly divergent Christensen 
subtraction term which had been
overlooked in the earlier calculations \cite{Jensen89}. 

\section{Effective Potential}

 Further insight into the nature of the modified black hole metric may
be gained by studying the motion of test particles in the vicinity of
the horizon. To this end, we compute the effective potential of the
perturbed black hole, as this completely characterizes the motion of both massless and massive test particles. 

We present a derivation for the effective potential for point
particles moving in a general static and spherically symmetric
background based on a Hamilton-Jacobi approach. The line element and
metric of such background spacetimes can always be cast in the form
\begin{equation}
ds^2 = g_{tt}(r)dt^2 + g_{rr}(r)dr^2 + r^2(d\theta^2 + \sin^2 \theta
d\phi^2).
\end{equation}
The trajectory of a point particle of mass $m$ moving in this
background
can be obtained from the Hamilton-Jacobi equation
\begin{equation}
g^{\mu \nu}\frac{\partial S}{\partial x^{\mu}}
\frac{\partial S}{\partial x^{\nu}} + m^2 = 0,
\end{equation}
where $S$ denotes the action (not to be confused with the entropy) of the
particle. As in every spherically symmetric field of force, the motion
occurs in a fixed plane passing through the origin; we take this plane
coincident with the slice defined by $\theta = \pi/2$ without any loss
of generality. Then, expanding out (51) gives the differential equation
for $S$,
\begin{equation}
g^{tt}\left( \frac{\partial S}{\partial t} \right)^2 + g^{rr}
\left( \frac{\partial S}{\partial r} \right)^2 + \frac{1}{r^2}
\left( \frac{\partial S}{\partial \phi} \right)^2 + m^2 = 0.
\end{equation}
By the general procedure for solving the Hamilton-Jacobi equation, we
look for an $S$ of the form \cite{Landau76} 
\begin{equation}
S(t,r,\phi) = -Et + L\phi + S_r(r),
\end{equation}
with constant energy $E$ and angular momentum $L$. Substituting this
ansatz into (52) gives an equation for $S_r$ which may be integrated
immediately to yield 
\begin{equation}
S_r(r) = \int^r (-g^{tt}\, g_{rr})^{1/2}\,
\left[ E^2 + g_{tt}(m^2 + \frac{L^2}{r^2}) \right]^{1/2} \, dr +
\delta_r,
\end{equation}
where $\delta_r$ is an arbitrary additive phase constant. The
dependence $r = r(t)$ for the radial coordinate of the particle is
given by
\begin{equation}
 \frac{\partial S}{\partial E} = \gamma = {\rm constant},
\end{equation}
or,
\begin{equation}
t = -\gamma + \int^r \frac{ (-g^{tt}g_{rr})^{1/2} E}{\left[
 E^2 + g_{tt}(m^2 + \frac{L^2}{r^2}) \right]^{1/2} }.
\end{equation}
This can be cast in terms of a differential equation for $r$ as
\begin{equation}
\left( \frac{dr(t)}{dt} \right) = (-g_{tt}g^{rr})^{1/2}\frac{1}{E}
\left[ E^2 + g_{tt}(m^2 + \frac{L^2}{r^2}) \right]^{1/2},
\end{equation}
which governs the radii of allowed orbits of particles moving in the
gravitational field represented by (22). Hence the function
\begin{equation}
V(r) \equiv -g_{tt} \, \left(m^2 + \frac{L^2}{r^2} \right),
\end{equation}
plays the role of the effective potential energy in the sense that the
condition $E^2 > V(r)$ determines the admissable range of the
particle's motion.
Identifying $g_{tt}$ from the solution (22) yields $V(r)$ for the
semiclassical black hole:
\begin{eqnarray}
V(w) &=& (1 - w) \left[ 1 + \epsilon(2{\bar \rho}(w) - w(1 -
w)^{-1}\mu(w) ) \right] \left(m^2 + \frac{L^2}{4M^2} w^2 \right) \\ \nonumber
     &=& V_o(w) \left( 1 + \epsilon V_1(w) \right),
\end{eqnarray}
and reduces, as it must, to the classical Schwarzschild potential as
$\epsilon \rightarrow 0$.
The relative correction $V_1$ depends on the cavity radius $r_o$ and
on $r < r_o$. For purposes of illustration, we consider the asymptotic radius
$r_o = 260M$, which is the maximum radius for which the metric 
perturbations (call them $\delta$) remain uniformly small: that is, 
$\delta < 1$ for
$2M \leq r < r_o$, as discussed at length in Ref \cite{HKY94}. 
(The perturbatively valid
domains for the spin-1 and spin-2 back-reactions are identical, by virtue of the fact that ${\hat b_1} = {\hat b_2}$.) A graphical
analysis of $V_1$ shows that the relative correction is large and negative for all $r \geq 2M$ and is insensitive to the particular value of $\hat c_2$ used to calculate it. In Fig. 2 we display $V_1(w)$ taking $\hat c_2 = -1400$. 
A glance at this Figure shows that $V_1 \sim -1.8$, so that a small amount of back-reaction, say $\epsilon = 0.1$, would lead to roughly a $20\%$ decrease in the effective potential. A decreased potential, in turn, 
implies an increase in the black hole capture cross section, as 
discussed in Ref. \cite{HKY94}.

\section{Discussion}

The solution of the lowest-order back-reaction of a conformal spin-2 field on a Schwarzschild black hole has been calculated based on the approximate stress energy tensor of Frolov and Zel'nikov. The new equilibrium metric has been found to order $O(\hbar)$.
By calculating the thermodynamical entropy by which the spin-2 field modifies the Bekenstein-Hawking entropy, we have been able to put an upper limit
on one of the constants $({\hat c_2})$ which parametrizes
this approximate stress-energy tensor. This is done by demanding that the fractional correction to the entropy be a positive and monotone increasing function of $r$, over the entire domain of perturbative validity of the solution.
This has led to an upper bound for $\hat c_2$.
( Conditions that general thermal stress-energy tensors must satisfy in order that $\Delta S \geq 0$ have been established recently by Zaslavskii \cite{Zas93}).
This bound translates physically into 
a {\em lower} limit for the magnitude of the quantum fluctuations of the spin-2 field at the black hole horizon.
By demanding that the metric perturbations be small over the range in which the perturbative solution is valid, we have also estimated a lower bound for
$\hat c_2$, though this particular bound depends on the cavity radius
$r_o$ and on the mass ratio $\epsilon = (M_{pl}/M)^2 < 1$. Nevertheless, these
limits translate, via the boundary condition (13), into corresponding bounds on the magnitude of the spin-2 field fluctuations at the black hole horizon.
 In the context of the spin-hierarchy of conformal field back-reaction, the spin-2 case is clearly the most important, giving by far, the largest correction. 
This is manifested in the sequence of 
spin-dependent horizon-energy densities and is exposed 
in a striking way in the relative correction to the black hole effective potential. 
It is hoped that these calculations may shed light on some of the gross features characterizing the graviton back-reaction. Certainly, the marked increase in the magnitude of the fluctuations in going from low spins to spin=2 is a 
feature we expect to persist in the graviton case, as well as the relative
increase in the entropy correction.

\vspace{.5cm}
\noindent
{\bf Acknowledgement}

One of us (DH) thanks Juan P\'erez-Mercader, Alvaro Dom\'inguez and Thomas
Buchert for useful comments.

\vfill\eject
\vspace{2cm}

\noindent
\underline{\bf Figure Captions}:

\noindent
Figure 1: Entropy correction $(\frac{3K}{4\pi}) \Delta S(w)$ due to 
the spin-2 back-reaction, for various values of $\hat c_2$.

\vspace{.6cm}

\noindent
Figure 2: Entropy corrections $(\frac{3K}{4\pi}) \Delta S(w)$ arising from 
spin-$0,\frac{1}{2},1$ back-reactions.

\vspace{.6cm}

\noindent
Figure 3: Relative correction $V_1(w)$ to the black hole effective potential.

\end{document}